\documentclass[aps,prb,twocolumn,superscriptaddress,floatfix,showpacs]{revtex4}
\usepackage{graphicx}
\usepackage{amsmath}
\usepackage{graphicx}

\begin{document}

\title{Band-mixing-mediated Andreev reflection of semiconductor holes}

\author{David Futterer}
\affiliation{Theoretische Physik, Universit\"at Duisburg-Essen and CeNIDE, 47048 Duisburg, Germany}
 
\author{Michele Governale}
\affiliation{School of Chemical and Physical Sciences and MacDiarmid Institute for Advanced Materials
and Nanotechnology, Victoria University of Wellington, PO Box 600, Wellington 6140, New Zealand}

\author{Ulrich Z\"ulicke}
\affiliation{School of Chemical and Physical Sciences and MacDiarmid Institute for Advanced Materials
and Nanotechnology, Victoria University of Wellington, PO Box 600, Wellington 6140, New Zealand}

\author{J\"urgen K\"onig}
\affiliation{Theoretische Physik, Universit\"at Duisburg-Essen and CeNIDE, 47048 Duisburg, Germany}

\date{\today}

\begin{abstract}
We have investigated Andreev-reflection processes occurring at a clean interface between a
$p$-type semiconductor and a conventional superconductor. Our calculations are performed
within a generalized Bogoliubov-de~Gennes formalism where the details of the semiconductor
band structure are described by a $6\times 6$ Kane model. It is found that Andreev reflection of
light-hole and heavy-hole valence-band carriers is generally possible and that the two
valence-band hole types can be converted into each other in the process. The normal-reflection and
Andreev-reflection amplitudes depend strongly on the semiconductor's carrier concentration
and on the angle of injection. In the special case of perpendicular incidence, Andreev
reflection of heavy holes does not occur. Moreover, we find conversion-less Andreev reflection to be impossible
above some critical angle, and another critical angle exists above which
the conversion of a heavy hole into a light hole cannot occur.
\end{abstract}


\maketitle

\section{Introduction}

Mesoscopic superconductivity has developed strongly over recent years.\cite{hekkingreview,
beenreview, mesoeltrans, lambertreview, schaepersbook, takayanagi_symp} Starting from
the early theoretical studies of superconductor--normal-metal (S-N) interfaces,\cite{andreev64,
degennes63} the interplay of pair correlations and quantum transport in phase-coherent
conductors has attracted a lot of interest.\cite{hekkingreview, beenreview, mesoeltrans,
lambertreview} As the charge carriers' mean free path can be much longer in semiconductors
than it typically is in metals, hybrid structures of semiconductor materials are ideal for
investigating the regime of ballistic transport.\cite{takayanagi85,schaepersbook} Most recently,
opportunities for realizing quantum-logical circuits and investigating fundamentals of quantum
physics in these systems have been explored.\cite{takayanagi_symp}

In most previous studies, the band electrons in the normal-conducting part of S-N hybrid
systems were simple in the sense that their properties could be modeled using quantum
states of free spin-1/2 particles. In the fundamentally interesting and practically
relevant\cite{bagraev09} situation where the normal carriers are from the valence band,
their electronic and spin properties are much richer.\cite{yu10, winkler03} States in the
uppermost valence bands of common semiconductor materials carry a spin-3/2 degree
of freedom and also exhibit a strong coupling between this larger spin and their orbital
motion. In our work presented here, we address the question of how these peculiar
features that have been seen to result in interesting mesoscopic-transport
effects\cite{lau98, papadakis00, danneau06, grbic07, koduvayur08, quay10} will affect
the physical properties of $p$-type semiconductor-superconductor hybrid systems.

Many of the interesting phenomena exhibited by S-N structures are fundamentally
due to the process of Andreev reflection,\cite{andreev64,degennes63} which is the
conversion of a charge-carrier incident on the interface from the normal side into its
charge-conjugated and time-reversed copy. This counterintuitive effect fundamentally
results from the fact that the two electrons forming a Cooper pair in the superconducting
condensate are from time-reversed states.\cite{deGennes} A superconductor in close
proximity to a normal conductor induces pair correlations between such states also on
the normal side of the hybrid system. As a result, a charge carrier with energy below
the gap for quasiparticle excitations in the superconductor can, upon incidence on the
S-N interface, combine with its appropriate partner to enter the superconducting side as
a Cooper pair. In the process, the normal conductor is left with a missing carrier, usually
referred to as a ``hole," that has all the attributes of the time-reversed partner of the incident
particle. As this hole is really a quasiparticle excitation of the Fermi sea of nearly-free
band electrons in the normal conductor, we avoid this nomenclature here and reserve
the term hole to always refer to a state in the valence band of the semiconductor
material making up the normal-conducting part of the hybrid structure.

Based on the Bogoliubov-de~Gennes formalism,\cite{deGennes} a theory for scattering
at nonideal S-N interfaces was developed by Blonder \textit{et al.} (BTK)
\cite{blonder82}. Later works have generalized this approach to describe the oblique
incidence of the charge carrier from the normal side\cite{raedt94, chaudhuri95, kupka97}
and to discuss the case of small values of Fermi energies typically realized in
semiconductors.\cite{mortensen99} It turns out that a finite angle of incidence
(measured with respect to the interface normal) reduces the probability of Andreev reflection,
and a critical angle exists above which no Andreev reflection is possible in a semiconductor.
The BTK model has also been adapted to situations without spin-rotational symmetry,
e.g., when the normal-conducting side of the hybrid system is ferromagnetic.\cite{jong95,
taddei01, xia02, eschrig09, grein10, kupferschmidt11,annunziata11} In the extreme case of a half-metallic
ferromagnet where only one spin-polarized band contributes to transport and pairing seems
to be impossible, spin-flip processes still enable Andreev reflections.\cite{eschrig09,
kupferschmidt11}

In the present work, we incorporate a $6\times 6$ Kane-type Hamiltonian\cite{winkler03}
into the Boguliubov-de~Gennes theory to model a hybrid $p$-type
semiconductor-superconductor structure. States in the lowest conduction and uppermost
(heavy-hole and light-hole) valence bands are included, as is the coupling between them.
We focus on the situation where the chemical potential lies in the valence band of the
semiconducting side and calculate the normal and Andreev-reflection probabilities when
either light holes or heavy holes are incident at an angle on the interface. In general, states
from the superconductor's conduction band will be incompatible with those from the
semiconductor's valence band due to their different orbital character, and no direct coupling
will be possible. Nevertheless, we find that the mixing between valence- and
conduction-band states in the semiconductor mediates a coupling to the superconducting
pair potential and thus enables Andreev reflection of holes. Even the valence-band states
with spin projection $\pm 3/2$ (heavy holes) can have a finite probability to be
Andreev reflected, even though the pair potential in the superconductor is between states
having spin projection $\pm1/2$. The wave-vector dependence of band mixing is reflected
in the variation of the Andreev-reflection amplitudes as a function of the holes' angle of
incidence onto the S-N interface.

The remainder of this paper is organized as follows. We introduce our model for a
$p$-type--semi\-con\-duc\-tor-super\-conductor hybrid structure and discuss its relevant physical
parameters in Sec.~\ref{sec:model}. Results for normal and Andreev-reflection
probabilities for different scenarios of incident heavy-hole and light-hole carriers are presented
in Sec.~\ref{sec:results}. A summary and conclusions of our work are given in Sec.~\ref{sec:concl}.

\section{Model}
\label{sec:model}

We consider a hybrid $p$-type semiconductor-superconductor structure with an ideal interface.
To calculate the transport properties of the system, we solve the Bogoliubov-De Gennes equation
with the single-particle Hamiltonians on the normal conducting and superconducting side formulated in Nambu space.
In order to avoid confusions with valence-band carriers, we do not use the term ``Nambu-hole.'' Instead, we will address the corresponding states as ``time-reversed'' states, indicated by a tilde, so that solely valence-band carriers are referred to as holes.
The relevant Hamiltonians will be invariant under time reversal, $\hat{H}=\Theta \hat{H} \Theta^{-1}$, where $\Theta$ is the time-reversal operator.

We model the semiconductor using a 6 $\times$ 6 Kane-Hamiltonian \cite{winkler03} within the spherical approximation in the basis of the $k=0$-band-edge states $\{ \vert \frac{1}{2} \left. \frac{1}{2} \right. \rangle_{c},\vert \frac{1}{2} \left. -\frac{1}{2} \right. \rangle_{c},\vert \frac{3}{2} \left. \frac{3}{2} \right. \rangle_{v},\vert \frac{3}{2} \left. \frac{1}{2} \right. \rangle_{v},\vert \frac{3}{2} \left. -\frac{1}{2} \right. \rangle_{v},\vert \frac{3}{2} \left. -\frac{3}{2} \right. \rangle_{v} \}$, representing conduction electrons, heavy holes, and light holes, respectively, and the corresponding time-reversed states, so that we get
\begin{equation}
\label{KaneHamiltonian}
H^N=
\begin{pmatrix}
\begin{matrix}
H_{0}^{6c 6c} & H_{0}^{6c 8v}\\
H_{0}^{8v 6c}&H_{0}^{8v 8v} 
\end{matrix}
& \mathbf{0}\\
\mathbf{0} & 
\begin{matrix}
-H_{0}^{6c 6c} & -H_{0}^{6c 8v}\\
-H_{0}^{8v 6c}&-H_{0}^{8v 8v} 
\end{matrix}
\end{pmatrix},
\end{equation}
with
\begin{subequations}
\label{HN}
\begin{equation}
\label{H6c6c}
H_{0}^{6c 6c}=
\begin{pmatrix}
\frac{\hbar^2 k^2}{2 m'}+E_{F}^N+E_0 & 0\\
0 &\frac{\hbar^2 k^2}{2 m'}+ E_{F}^N+E_0
\end{pmatrix},
\end{equation}
\small
\begin{widetext}
\begin{equation}
\label{H8v8v}
H_0^{8v 8v}=\begin{pmatrix}
\begin{matrix}
-\frac{\hbar^2 (k_x^2+k_y^2)}{2 m_0}
(\gamma'_{1}+\gamma'_2)\\
-\frac{\hbar^2 k_z^2}{2 m_0} (\gamma'_1-2 \gamma'_2)+E_F^N
\end{matrix}
&
2 \sqrt{3} \frac{\hbar^2 k_{-}k_z}{2 m_0} \gamma'_3
&
\begin{matrix}
\sqrt{3} \frac{\hbar^2 \hat{K}}{2 m_0} \gamma'_2\\
-2 i \sqrt{3} \frac{\hbar^2 k_x k_y}{2 m_0} \gamma'_3
\end{matrix}
&
0\\
\\
2 \sqrt{3} \frac{\hbar^2 k_+ k_z}{2 m_0} \gamma'_3
&
\begin{matrix}
-\frac{\hbar^2 (k_x^2+k_y^2)}{2 m_0} (\gamma'_1-\gamma'_2)\\
-\frac{\hbar^2 k_z^2}{2 m_0}(\gamma'_1+2 \gamma'_2)+E_F^N
\end{matrix}
&
0
&
\begin{matrix}
\sqrt{3} \frac{\hbar^2 \hat{K}}{2 m_0}\gamma'_2\\
-2 i \sqrt{3} \frac{\hbar^2 k_x k_y}{2 m_0} \gamma'_3
\end{matrix}
\\
\\
\begin{matrix}
\sqrt{3} \frac{\hbar^2 \hat{K}}{2 m_0} \gamma'_2\\
+2 i \sqrt{3} \frac{\hbar^2 k_x k_y}{2 m_0} \gamma'_3
\end{matrix}
&
0
&
\begin{matrix}
-\frac{\hbar^2 (k_x^2+k_y^2)}{2 m_0} (\gamma'_1-\gamma'_2)\\
-\frac{\hbar^2 k_z^2}{2 m_0}(\gamma'_1+2 \gamma'_2)+E_F^N
\end{matrix}
&
-2 \sqrt{3} \frac{\hbar^2 k_- k_z}{2 m_0} \gamma'_3\\
\\
0
&
\begin{matrix}
\sqrt{3} \frac{\hbar^2 \hat{K}}{2 m_0}\gamma'_2\\
+2 i \sqrt{3} \frac{\hbar^2 k_x k_y}{2 m_0} \gamma'_3
\end{matrix}
&
-2 \sqrt{3} \frac{\hbar^2 k_+ k_z}{2 m_0} \gamma'_3
&
\begin{matrix}
-\frac{\hbar^2 (k_x^2+k_y^2)}{2 m_0}(\gamma'_1+\gamma'_2)\\
-\frac{\hbar^2 k_z^2}{2 m_0}(\gamma'_1-2 \gamma'_2) +E_F^N
\end{matrix}
\end{pmatrix},
\end{equation}
\end{widetext}
\normalsize
\begin{eqnarray}
&&H_0^{6c 8v}=(H_0^{8v 6c})^{\dagger} \nonumber \\ && =
\begin{pmatrix}
-\frac{1}{\sqrt{2}} P k_+ & \sqrt{\frac{2}{3}} P k_z & \frac{1}{\sqrt{6}} P k_- & 0\\
0 & -\frac{1}{\sqrt{6}} P k_+ & \sqrt{\frac{2}{3}} P k_z & \frac{1}{\sqrt{2}} P k_-
\end{pmatrix} \label{coupling},
\end{eqnarray}
\end{subequations}
and $\mathbf{0}$ being the zero matrix of the appropriate dimensions.
The spherical approximation implies that the terms arising from bulk inversion asymmetry can be neglected and that $\gamma'_{1}\rightarrow  \gamma_1-\frac{1}{3}\frac{2 m_0}{\hbar^2}\frac{P^2}{E_0}$ and $\gamma'_{2,3} \rightarrow \frac{2 \gamma_{2}+3 \gamma_{3}}{5}-\frac{1}{6}\frac{2 m_0}{\hbar^2}\frac{P^2}{E_0}$. We have used the abbreviations $m'=m_0 \left( \frac{m_0}{m^*}-\frac{2}{3} \frac{2 m_0}{\hbar^2} \frac{P^2}{E_0} \right)^{-1}$, where $m^*$ is the effective mass of conduction band electrons, $k^2=k_x^2+k_y^2+k_z^2$, $\, k_{\pm}=k_x \pm i k_y$, and $\hat{K}=k_x^2-k_y^2$.
The Fermi energy of the semiconductor is $E^{N}_{F}$, $E_{0}$ is the energy gap between the conduction band and valence bands, $P$ is the coupling parameter between the conduction band and the valence band, and $\gamma_{1,2,3}$ are parameters generating the effective masses in the valence band.
\begin{figure}
\includegraphics[width=0.9\columnwidth]{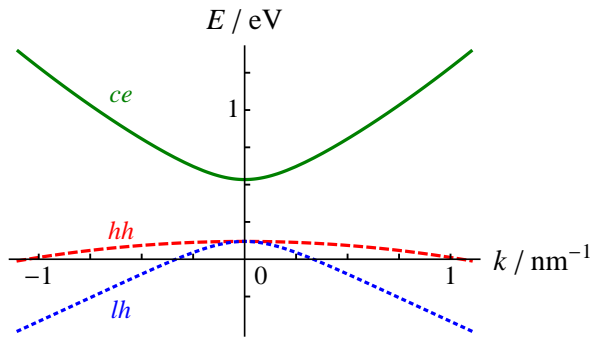}
\caption{(Color online)
Schematic dispersions of conduction electrons (ce), heavy holes (hh) and light holes (lh).
\label{band-structure}}
\end{figure}
Figure \ref{band-structure} schematically shows the dispersion resulting from the $6\times 6$ Kane-Hamiltonian, exemplarily calculated with the parameters of InAs, which will be discussed in more detail in Sec.~\ref{sec:results}, and a Fermi energy of $E_F^N=110$ meV corresponding to a carrier concentration $n$ of $n=10^{20}\mathrm{cm^{-3}}$.

In the superconductor, we assume the gap between the conduction band and the valence bands to be very large, so that valence-band states of the superconductor are irrelevant. In order to be able to match the wave function in the superconductor with the wave function in the semiconductor, we write $H^S$ also in the $12\times 12$ basis of the $k=0$-band-edge states and shift the valence bands in the superconductor to minus infinity. Then the Hamiltonian $H^{S}$ of the superconductor is given by
\begin{equation}
H^S=
\begin{pmatrix}
\label{HS}
H_0^S & \mathbf{0} & \Delta \cdot \mathbf{I}_{2 \times 2} & \mathbf{0}\\
\mathbf{0} & -\infty \cdot \mathbf{I}_{4 \times 4} & \mathbf{0} & \mathbf{0}\\
\Delta^* \cdot \mathbf{I}_{2 \times 2}& \mathbf{0} & - H_0^S & \mathbf{0}\\
\mathbf{0} & \mathbf{0} & \mathbf{0} & \infty \cdot \mathbf{I}_{4 \times 4}
\end{pmatrix},
\end{equation}
where
\begin{equation}
\label{H0S}
H_{0}^{S}=
\begin{pmatrix}
\frac{\hbar^2 k^{2}}{2 m_{S}}-E_{F}^{S} & 0\\
0 & \frac{\hbar^2 k^{2}}{2 m_{S}} -E_{F}^{S}
\end{pmatrix},
\end{equation}
with $\mathbf{I}_{n\times n}$ being the identity matrix in $n$ dimensions, $m_{S}$ being the effective mass of the superconductor, and $E_F^S$ being its Fermi energy. Without loss of generality, we choose the superconducting order parameter $\Delta=\Delta_0$ to be real. It follows from Eq.~(\ref{HS}) that only states of the semiconductor with nonzero $| \frac{1}{2} \left.\pm \frac{1}{2} \right. \rangle_{c}$ component couple to the superconductor.
In the following, we consider either the injection of a light hole (lh) or the injection of a heavy hole (hh) from the semiconducting side.
For oblique reflections, it is sufficient to consider all particles to move in a plane, which we choose to be the $x$-$y$ plane.
This choice block diagonalizes the Kane-Hamiltonian and thus reduces the full Bogoliubov-de Gennes Hamiltonian from $12 \times 12$ to $6 \times 6$.
We assume the N-S interface to be in the $y$-$z$ plane at $x=0$.
For an injected light hole $\xi=lh$ (heavy hole $\xi=hh$) in the semiconductor, we make the following ansatz:
\begin{eqnarray}
\psi_{\xi}(\mathbf{x})=&&\frac{1}{\sqrt{|v_{\xi}(k_{\xi}^{i})|}} \mathbf{u}_{\xi}( k_{\xi}^{i}) e^{i k_{\xi,\perp}^{i} x_{\perp}+i k_{\parallel} x_{\parallel}}\nonumber\\
&&+\sum_{\chi} \frac{r_{\chi/\xi}}{\sqrt{|v_{\chi}(k_{\chi}^{r})|}} \mathbf{u}_{\chi}( k_{\chi}^{r}) e^{i k_{\chi,\perp}^{r} x_{\perp}+i k_{\parallel} x_{\parallel}},
\end{eqnarray}
where $x_{\perp}$ ($x_{\parallel}$) is the component perpendicular (parallel) to the junction, $\mathbf{u}_{\chi}(k)$ is the eigenvector corresponding to state $\chi$ and momentum $k$ and $v_{\chi}(k)=\mathbf{u}_{\chi}(k)^T \hat{v}_{\perp} \mathbf{u}_{\chi}(k)$, with $\hat{v}_{\perp}=\frac{i}{\hbar} \left[ H,\mathbf{x_{\perp}} \right]$, denotes the velocity perpendicular to the junction. The reflection coefficient describing the reflection amplitude from state $\xi$ into state $\chi$ is labeled $r_{\chi/\xi}$. The index $\chi \in \{ ce,hh,lh,\tilde{ce},\tilde{hh},\tilde{lh} \}$ denotes a combination of the band (conduction band, heavy-holes band, light-holes band) and the Nambu state (non-time-reversed, time-reversed). Due to the fact that in scattering processes the momentum parallel to the interface needs to be conserved, all parts of the wave function have the same momentum parallel to the scattering interface $k_{\parallel}$. With $k_{\parallel}$ and $k_{\xi,\perp}^{i}$ ($k_{\chi,\perp}^{r}$), the angle $\theta$ of the injected (reflected) particle is determined, where $\theta=0$ corresponds to the case of normal incidence.
Note that we explicitly allow for a conversion between conduction electrons, light holes, and heavy holes, i.e., we allow for light holes to be normal reflected as heavy holes and conduction electrons, and Andreev reflected as heavy holes and conduction electrons; and, analogously, we allow for heavy holes to be normal reflected as light holes and conduction electrons, and Andreev reflected as light holes and conduction electrons. Since the semiconductor's conduction band lies above the Fermi energy, only evanescent conduction-electron modes exist. But nevertheless, these modes are important for matching the wave functions at the boundary.

We restrict ourselves to excitation energies inside the superconducting gap, $|E| < \Delta_{0}$, which implies that only evanescent quasiparticle-wave functions exist in the superconductor.
For the wave function in the superconductor, we set

\begin{eqnarray}
\psi_{S}(\mathbf{x})=&&\frac{c_{ce}}{\sqrt{\mathrm{Re}[q_{ce}]}}
\begin{pmatrix}
\gamma^{*}\\ 
0\\
0\\
\gamma\\
0\\
0
\end{pmatrix}
e^{i q_{ce,\perp} x_{\perp}+i k_{\parallel} x_{\parallel}}+\nonumber\\
&& \frac{c_{\tilde{ce}}}{\sqrt{\mathrm{Re}[q_{\tilde{ce}}]}}
\begin{pmatrix}
\gamma\\
0\\
0\\
\gamma^{*}\\
0\\
0
\end{pmatrix}
e^{i q_{\tilde{ce},\perp} x_{\perp}+i k_{\parallel} x_{\parallel}},
\end{eqnarray}
with
\begin{equation}
\gamma=\exp\left[ -\frac{i}{2} \arccos \left( \frac{E}{\Delta_{0}} \right) \right],
\end{equation}
where $c_{ce}$ and $c_{\tilde{ce}}$ are transmission coefficients and $q_{ce}$ ($q_{\tilde{ce}}$) is the complex wave vector of the (time-reversed) evanescent quasiparticle wave function.

At the junction, the wave function and the velocity need to be continuous:
\begin{eqnarray}
\psi_{N}(x_{\perp}=0)&=&\psi_{S}(x_{\perp}=0)\\
\hat{v}_{\perp} \psi_{N}(x_{\perp}=0) &=&\hat{v}_{\perp} \psi_{S}(x_{\perp}=0),
\end{eqnarray}
with $\psi_{N}$ being $\psi_{lh}$ or $\psi_{hh}$, respectively.\\

\section{Results}
\label{sec:results}

The results shown in this section have been calculated for InAs-Al. InAs is a commonly used material that meets the requirements of a large mixing between conduction band and valence bands, described by a large value of $P$ and a small energy gap $E_0$ between conduction band and valence bands, as well as a large spin-orbit coupling so that the spin split-off band can be neglected, and Al is often used by experimentalists as a superconducting material. We use the band structure parameters for InAs given in Ref.~\onlinecite{winkler03}, which are $E_0=0.418 \; \mathrm{ eV}, \; P=9.197 \; \mathrm{ eV \AA},m^{*}=0.0229 \; m_{0},\gamma_{1}=20.40,\gamma_{2}=8.30$, and $\gamma_{3}=9.10$. Typical carrier concentrations $n$ of $p$-type InAs range from about $10^{16} \; \mathrm{cm}^{-3}$ up to about $10^{20} \; \mathrm{cm}^{-3}$. This corresponds to Fermi energies ranging from about $0.2$ meV up to about $120$ meV.
For Al, we set $m_{S}= m_0$ and $E_{F}^{S}=11.63$ eV.

\begin{figure}
\includegraphics[width=0.8\columnwidth,angle=270]{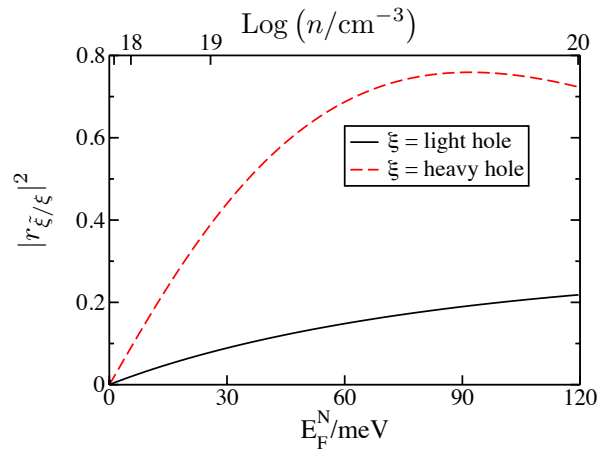}
\caption{(Color online)
Plot of the conversionless Andreev-reflection probabilities as a function of the Fermi energy of the semiconductor $E_F^N$ or the corresponding carrier concentration of the semiconductor. The other parameters are $\Delta_{0}=0.1 \; \mathrm{meV}, \; \theta=\pi/8$, and $E=0$.
\label{AR-sequence}}
\end{figure}

It follows from Eqs.~(\ref{HN}) that all semiconductor states with $k=0$ are orthogonal to each other. This implies that the valence-band states at $k=0$ are orthogonal to the ($k=0$) conduction-band states $| \frac{1}{2} \left.\pm \frac{1}{2} \right. \rangle_{c}$. They are, therefore, decoupled from the superconductor and Andreev reflections are not possible. Only valence-band states with finite momentum $k$ can have a finite $| \frac{1}{2} \left.\pm \frac{1}{2} \right. \rangle_{c}$ component so that these states can participate in Andreev-reflection processes.
 At carrier concentrations in the range of $10^{16} \; \mathrm{cm}^{-3}$ in the semiconductor, the Fermi energy is smaller than $1$ meV causing the injected hole to have a small Fermi momentum. In this situation, Andreev-reflection probabilities are strongly suppressed; see Fig. \ref{AR-sequence}. At larger carrier concentrations, the Fermi energy is shifted away from the band edge causing the Fermi momentum to be increased, and the probability of Andreev reflection is finite; see Fig. \ref{AR-sequence}.

\begin{figure*}
\includegraphics[width=1.5\columnwidth,angle=270]{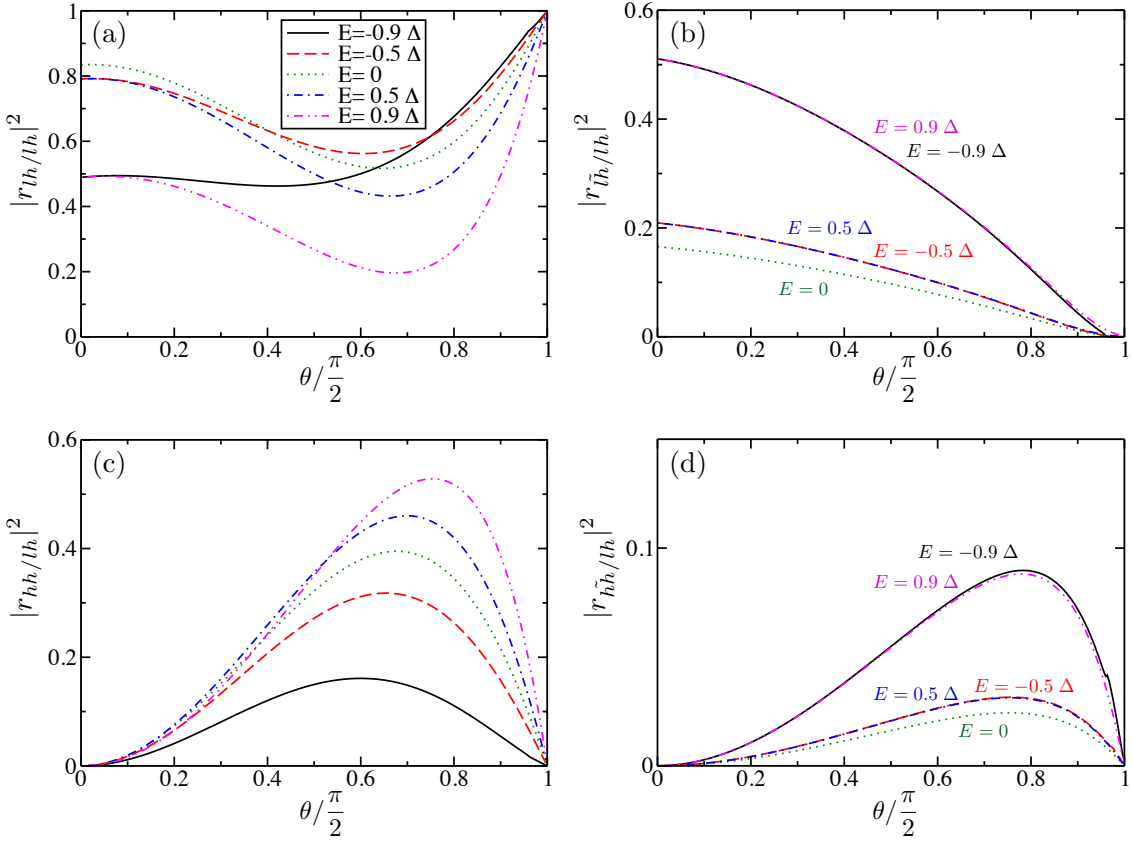}
\caption{(Color online)
Plot of the reflection probabilities for an injected light hole as a function of the injection angle $\theta$ for different excitation energies $E$. The other parameters are $\Delta_{0}=0.1$ meV and $E_{F}^{N}=53.6$ meV, which corresponds to $n=3 \times 10^{19}\mathrm{cm^{-3}}$.
\label{lh-10-to-19}}
\end{figure*}

\begin{figure*}
\includegraphics[width=1.5\columnwidth,angle=270]{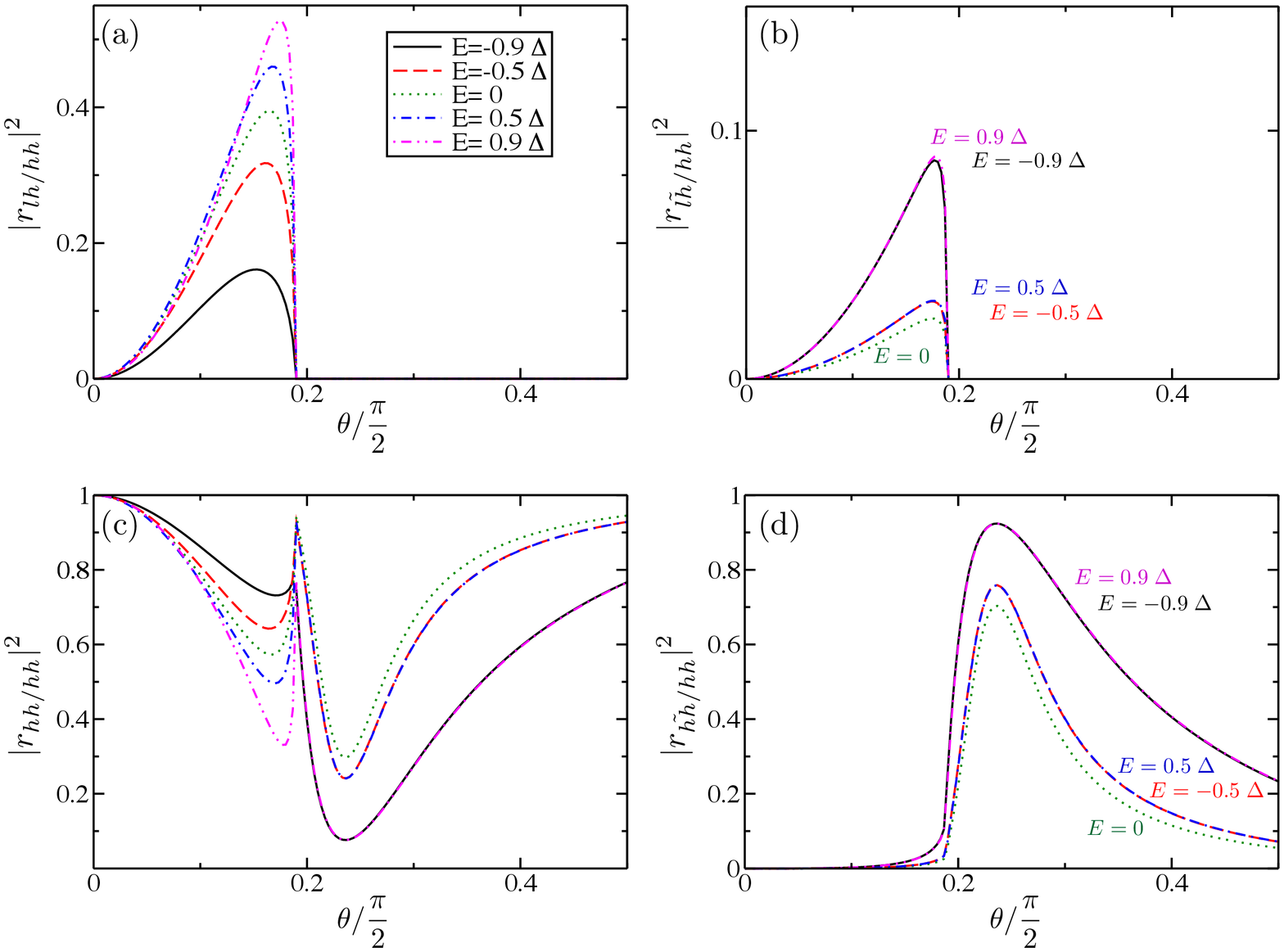}
\caption{(Color online)
Plot of the reflection probabilities for an injected heavy hole as a function of the injection angle $\theta$ for different excitation energies $E$. The other parameters are $\Delta_{0}=0.1$ meV and $E_{F}^{N}=53.6$ meV, which corresponds to $n=3 \times 10^{19}\mathrm{cm^{-3}}$.
\label{hh-10-to-19}}
\end{figure*}

Light holes and heavy holes are distinguished by the projection of their total angular momentum in the direction of motion. If a heavy hole is incident onto a scattering interface at a finite angle with the surface normal, the reflected state would have a different spin-quantization axis and would, therefore, be a mixture of heavy-hole and light-hole components. As a result, it is possible to convert heavy holes into light holes, and vice versa, in oblique scattering processes. This conversion also occurs in Andreev-reflection processes, so that heavy holes may also be Andreev reflected as light holes, and vice versa.
For normal incidence, heavy holes are decoupled from light holes and conduction electrons. This can be seen by setting $k_x$ and $k_y$ to zero in Eq.~(\ref{KaneHamiltonian}). For perpendicular (and, in principle, also parallel) incidence, the motion of incoming and reflected particles is collinear so that, in this case, a conversion is not possible and, additionally, for heavy holes, it is not possible to be Andreev reflected. 
This effect is independent of the semiconductor's carrier concentration and independent of the excitation energy of the incident hole. Figures \ref{lh-10-to-19} and \ref{hh-10-to-19} show an exemplary sequence of plots of the reflection probabilities of an injected light hole or injected heavy hole, respectively, as a function of the angle of injection for different excitation energies.
In general, the conversion between heavy holes and light holes via normal reflection and Andreev reflection is possible, but not in the limits of perpendicular incidence ($\theta=0$) or parallel incidence ($\theta=\pi/2$). For heavy holes, also, the probability for Andreev reflection without conversion vanishes in these limits, so that we get $|r_{hh/hh}|^2=1$ and $|r_{\chi/hh}|^2=0$, for $\chi \neq hh$. In contrast to Andreev-reflection probabilities of conduction electrons, which in general get reduced by increased angles of injection, we find that heavy holes require a nonzero angle of injection to be Andreev reflected.\\
For perpendicularly incident light holes, we are able to derive analytical results in the limit of the Andreev approximation, i.e., $|E| \ll E_{F}^{N,S}$ and $\Delta_{0} \ll E_{F}^{N,S}$. This is a reasonable assumption as long as the semiconductor is doped such that its Fermi energy is large compared to the pair potential $\Delta_0$. The Andreev approximation implies $q_{ce}\approx -q_{\tilde{ce}}$ as well as $k_{lh}^{i}\approx -k_{lh}^{r} \approx k_{\tilde{lh}}^{r}$ and $k_{ce}^{r} \approx k_{\tilde{ce}}^{r}$. 
In this limit, we find that the Andreev-reflection probability of light holes is of the BTK form,\cite{blonder82}
\begin{equation}
\label{BTK-result}
|r_{\tilde{lh}/lh}|^2=\frac{\Delta_0^2}{E^2+(\Delta_0^2-E^2)(1+2 Z^2)^2} \quad ,
\end{equation}
with all materials-specific quantities entering into a single interface parameter given by
\begin{widetext}
\begin{equation}
Z=\left[ \frac{\left( \frac{m_S}{m_0} k_{lh}^i-\frac{m'}{m_0} q_{ce} \right)^2 - \left( \frac{m_S}{m_0}+ \frac{\hbar^2 k_{lh}^i q_{ce}}{2 m_0} \frac{1}{E_F^N+E_0} \right)^2 (k_{ce}^r)^2}{4\frac{m_S}{m_0}k_{lh}^i q_{ce}\left( \frac{m'}{m_0}+\frac{\hbar^2 (k_{ce}^r)^2}{2 m_0} \frac{1}{E_F^N+E_0} \right)} \right]^{\frac{1}{2}}.
\end{equation}
\end{widetext}
Since no conversion between light holes and heavy holes occurs in the case of normal incidence, we get $|r_{lh/lh}|^2=1-|r_{\tilde{lh}/lh}|^{2}$.
For energies close to the superconducting gap, $|E| \rightarrow \Delta_{0}$, the probability for Andreev reflection approaches unity, $|r_{\tilde{lh}/lh}|^2 \rightarrow 1$, i.e., only Andreev reflections take place.

\begin{figure}
\includegraphics[width=0.67\columnwidth,bb=200 200 400 600]{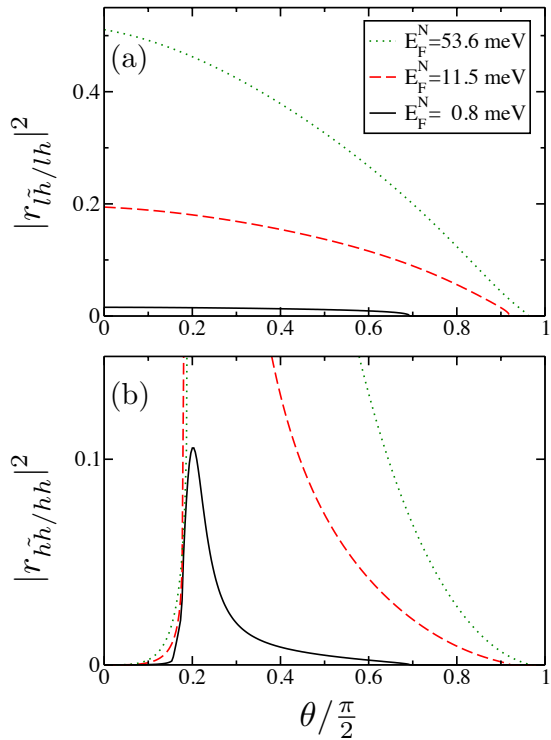}
\caption{(Color online)
Plot of the conversionless Andreev-reflection probabilities as a function of the injection angle $\theta$ for different Fermi energies of the semiconductor, where $E_F^N=53.6$ meV corresponds to $n=3 \times 10^{19}\mathrm{cm^{-3}}$, $E_F^N=11.5$ meV corresponds to $n=3 \times 10^{18}\mathrm{cm^{-3}}$, and $E_F^N=0.8$ meV corresponds to $n=5 \times 10^{16}\mathrm{cm^{-3}}$. The other parameters are $\Delta_{0}=0.1$ meV and $E=-0.9 \; \Delta$.
\label{critical-angle}}
\end{figure}

In addition to the above-discussed effects, we find two different types of critical angles. On the one hand, a critical angle occurs above which an injected hole cannot be Andreev reflected without conversion; see Fig. \ref{critical-angle}. Furthermore, for an incident heavy hole, there is a critical angle for reflections associated with a conversion to a light hole; see Figs. \ref{hh-10-to-19}(a) and \ref{hh-10-to-19}(b). Both types of critical angle have the same physical origin: the momentum component parallel to a planar interface needs to be conserved in the scattering process. If this parallel component of the incident particle is larger than the total momentum available at a given energy for a particular type of reflected particle, the associated process of reflection is not possible. In contrast to the critical angle of conduction electrons discussed by Mortensen \textit{et al.},\cite{mortensen99} the critical angle for conversionless Andreev reflection of holes occurs for negative excitation energies (as measured from the Fermi energy of the hole carriers). Due to the shape of the dispersion in the semiconductor (see Fig. \ref{band-structure}), an injected hole with excitation energy below the Fermi energy has a larger momentum than the corresponding time-reversed hole, thus, a critical angle exists at which the parallel component of the total momentum of the injected light hole (heavy hole) equals the total momentum of the time-reversed light hole (heavy hole):
\begin{equation}
\sin{\theta_{cl}^{lh(hh)}}=\frac{|k_{\tilde{lh}(\tilde{hh})}|}{|k_{lh(hh)}|}.
\end{equation}
This critical angle requires finite excitation energies (otherwise the injected hole and the reflected time-reversed hole have the same magnitude of the momentum) and is more pronounced for small Fermi energies because then the ratio of $|k_{\tilde{lh}(\tilde{hh})}|/|k_{lh(hh)}|$ becomes smaller. These considerations are in very good agreement with the plotted results (see Fig. \ref{critical-angle}): the critical angle is visible at negative excitation energies for light holes as well as for heavy holes, and has a smaller value for smaller Fermi energies.

The second type of critical angle can also be understood by looking at the semiconductor's dispersion, as shown in Fig. \ref{band-structure}. Close to the Fermi energy, a heavy hole has a much larger total momentum compared to a light hole of the same energy. Thus, a critical angle exists, at which the parallel component of the total momentum of an injected heavy hole equals the total momentum of the corresponding (time-reversed) light hole:
\begin{equation}
\sin{\theta_c^{lh(\tilde{lh})}}=\frac{|k_{lh(\tilde{lh})}|}{|k_{hh}|}.
\end{equation}
This critical angle is, to a good approximation, a constant function of the Fermi energy of the semiconductor and of the excitation energy of the injected hole. In Figs. \ref{hh-10-to-19}(a) and \ref{hh-10-to-19}(b), it is clearly seen that the probabilities for a heavy hole to be normal reflected or Andreev reflected as a light hole vanish
nearly independently of the excitation energy at some critical angle.\\

\section{Conclusions}
\label{sec:concl}

We have studied reflection of light holes and heavy holes at subgap energies from the interface of a $p$-type semiconductor with a conventional superconductor.
As a main result of this paper, we find that Andreev reflection of light holes as well as heavy holes is possible.
This is a consequence of spin-orbit coupling that mixes states of different angular momenta, i.e., spin is not a good quantum number anymore.
Due to this band mixing, the valance-band states are no longer purely described by a spin-$3/2$ degree of freedom with spin projection $\pm 3/2$ for heavy and $\pm 1/2$ for light holes.
Instead, they are linear combinations that contain, in general, a finite component of the spin-$1/2$ degrees of freedom that dominate the semiconductor's conduction band and couple to the superconductor.
It is this mixed-in component that enables Andreev reflection of the incident heavy or light hole at the interface.
Furthermore, the band mixing couples the different spin projections of the spin-$3/2$ degree of freedom.
This opens the possibility for conversion of heavy holes into light holes, and vice versa (both during normal and Andreev reflection).

The strength of the band mixing and, thus, the Andreev-reflection amplitude of light holes as well as heavy holes depends strongly on the angle of incidence and the Fermi energy (i.e., the carrier density) in the semiconductor.
In particular, we find the following:\\
(i) Light holes as well as heavy holes require a finite coupling to the conduction-band states to experience Andreev reflection, and this coupling can be increased by doping the semiconductor.\\
(ii) In the special case of perpendicular incidence, there is no coupling of heavy-hole states to light-hole or conduction-band states. Therefore, heavy holes can only be normal reflected as heavy holes, and the conversion between heavy holes and light holes is impossible.\\
(iii) Critical angles exist for conversionless Andreev reflection, and for the conversion of heavy holes into light holes and time-reversed light holes.

\acknowledgements
D.F. appreciates the hospitality of The Victoria University of Wellington. Financial support from DFG via KO 1987/5 and SFB 491 is gratefully acknowledged.

\end{document}